\documentclass[11pt,a4paper]{article}

\usepackage[margin=1in]{geometry}
\usepackage{amsmath,amssymb,amsthm}
\usepackage{mathtools}
\usepackage[numbers]{natbib}
\usepackage{hyperref}

\hypersetup{
    colorlinks=true,
    linkcolor=blue,
    citecolor=blue,
    urlcolor=blue
}

\theoremstyle{plain}
\newtheorem{theorem}{Theorem}

\newtheorem{lemma}{Lemma}[theorem]

\theoremstyle{definition}
\newtheorem{definition}[theorem]{Definition}

\newtheorem{remark}[theorem]{Remark}

\title{A metric on programs computing the same function}
\author{Konstantin Doubrovinski%
  \thanks{Email: \texttt{Konstantin.Doubrovinski@UTSouthwestern.edu}} \\
  \small University of Texas Southwestern Medical Center
}
\date{\today}

\begin{document}

\maketitle

\begin{abstract}
The question of proof equivalence (notably contemplated by David Hilbert for his 1900 list of problems but ultimately omitted) may be linked to the problem of program equivalence in computability theory. As was previously noted, identifying programs computing the same partial recursive function is too coarse, while demanding exact syntactic identity of program source code is overly strict.  In this work, we present an algorithmic approach within recursion theory to quantify the similarity between programs computing the same function. By considering the class of recursive endomorphisms (interpreters) and measuring their descriptive complexity, we construct an ultrametric distance on G\"odel numbers of a fixed partial recursive function. We prove that this metric is recursively invariant, non-trivial (i.e. unbounded), and satisfies the standard axioms of a metric space.
\end{abstract}

\section{Introduction}

Determining whether two mathematical proofs are ``essentially identical'' was famously contemplated by David Hilbert for inclusion in his 1900 Paris address. Under the Curry--Howard correspondence, propositions correspond to types and proofs correspond to terms or programs. Consequently, the proof-identity question naturally translates into a fundamental question in theoretical computer science: \emph{When do two programs represent the same algorithm?} \cite{buss2001prospects, blass2009two,yanofsky2011towards}.

In classical recursion theory, two standard equivalence relations present themselves:
\begin{enumerate}
    \item \textbf{Extensional equivalence:} Two programs are equivalent if and only if they define the same partial recursive function. This classification is excessively coarse. For instance, an optimal polynomial-time algorithm for finding a minimum spanning tree \cite{hopcroft2001introduction} and an exhaustive super-exponential brute-force search implement identical input-output behaviors, yet their algorithmic structures are distinct.
    \item \textbf{Syntactic identity:} Two programs are equivalent if and only if their codings are identical. This classification is excessively fine: appending comments or trivial whitespace modifications yields functionally identical implementations that intuitive algorithmic identity ought to equate.
\end{enumerate}

To bridge this gap, intermediate criteria have been explored, often relying on bisimulations, abstract state machines, or category-theoretic structures \cite{blass2009two,yanofsky2011towards}. Here, we explore an alternative, quantitative formulation grounded in algorithmic information theory \cite{li2008introduction,shen2017kolmogorov}. Rather than introducing a binary equivalence relation that risks either collapsing distinct algorithms or violating compositionality, we equip the fibers of a G\"odel numbering with a metric induced by chains of semantics-preserving total recursive transformations.

\section{Preliminaries and Definitions}

Our notation follows \cite{vereshchagin2003computable,rogers1987theory}. Let $\mathcal{P}^{(1)}$ denote the set of unary partial recursive functions from $\mathbb{N}$ to $\mathbb{N}$. We fix an acceptable G\"odel numbering $\{\phi_i\}_{i \in \mathbb{N}}$ of $\mathcal{P}^{(1)}$. Recall that a numbering is an acceptable G\"odel numbering if:
\begin{enumerate}
    \item Every partial recursive function $\psi \in \mathcal{P}^{(1)}$ appears in the enumeration; that is, $\exists i \in \mathbb{N}$ such that $\phi_i = \psi$.
    \item For every two-place partial recursive function $V(j, x)$, there exists a total recursive function $s \colon \mathbb{N} \to \mathbb{N}$ such that $V(j, x) = \phi_{s(j)}(x)$ for all $j, x \in \mathbb{N}$.
\end{enumerate}

Let $\phi_i \cong \phi_j$ denote extensional equality; i.e., $\forall x \in \mathbb{N}$, $\phi_i(x) \downarrow \iff \phi_j(x) \downarrow$, and whenever both halt, $\phi_i(x) = \phi_j(x)$. For any partial recursive function $f \in \mathcal{P}^{(1)}$, we denote its semantic fiber by
\begin{equation}
    \mathcal{F}_f = \{ i \in \mathbb{N} \mid \phi_i \cong f \}.
\end{equation}
By Kleene's recursion theorem, each fiber $\mathcal{F}_f$ is infinite \cite{rogers1987theory}.

\begin{definition}[Semantics-Preserving Interpreter]
A total recursive function $s \colon \mathbb{N} \to \mathbb{N}$ is called a \emph{semantics-preserving interpreter} (or an \emph{interpreter}) on the numbering $\{\phi_i\}$ if for all $i \in \mathbb{N}$,
\begin{equation}
    \phi_{s(i)} \cong \phi_i.
\end{equation}
Let $\mathcal{S}$ denote the set of all such interpreters.
\end{definition}

Let $K(\cdot)$ denote plain Kolmogorov complexity relative to a fixed universal Turing machine \cite{shen2017kolmogorov}. For a total recursive function $s \in \mathcal{S}$, $K(s)$ is understood as the least Kolmogorov complexity of $i$ with $\phi_i$ computing $s$.

\begin{definition}[Transformation Metric on Semantic Fibers]
\label{def:metric}
Let $i, j \in \mathcal{F}_f$ for some $f \in \mathcal{P}^{(1)}$. We define the distance $d(i, j)$ by
\begin{equation}
    d(i, j) = \inf \left\{ \max_{1 \le m \le n} K(s_m) \;\middle|\; n \in \mathbb{N}, \; s_1, \dots, s_n \in \mathcal{S}, \; s_n(\dots s_1(i) \dots) = j \right\}.
\end{equation}
\end{definition}

\section{Structural Properties of the Metric}

An apparent objection to this metric is that arbitrary pairs of extensionally equivalent programs might be connected via chains of elementary, low-complexity syntactic edits. To formalize this scenario, consider an interpreted syntax in which every program has the partitioned structure:
\begin{verbatim}
    b (code_1) S (code_2)
\end{verbatim}
where $b \in \{1, 2\}$ is an execution selector and \texttt{S} is a reserved separator symbol. If $b = 1$, the universal machine executes \texttt{code\_1} and treats \texttt{code\_2} as inactive comments; if $b = 2$, it executes \texttt{code\_2}. Under this scheme, one might attempt to deform any program $i$ into an extensionally equivalent program $j$ as follows:
\begin{enumerate}
    \item Initialize $b = 1$ with $\mathtt{code\_1} = i$, and clear the dormant partition $\mathtt{code\_2}$.
    \item Append the code of $j$ into $\mathtt{code\_2}$ character by character; because these modifications occur strictly within the inactive partition, each step constitutes a low-complexity operation that leaves the computed function unchanged.
    \item Toggle the selector bit $b$ from $1$ to $2$, activating $j$ and demoting $i$ to a comment.
\end{enumerate}
While steps (1) and (2) are admissible low-complexity operations, this sequence does not yield a valid path in our metric space because step (3) cannot be implemented by an interpreter $s \in \mathcal{S}$ of low descriptive complexity. By Definition~\ref{def:metric}, an interpreter must satisfy $\phi_{s(k)} \cong \phi_k$ uniformly for \emph{all} indices $k \in \mathbb{N}$. A global computable operator that toggles $b$ from $1$ to $2$ preserves semantics if and only if the active and dormant partitions compute the same function. Because extensional equivalence $\phi_u \cong \phi_v$ is $\Pi_2^0$-complete, an interpreter cannot verify this equivalence algorithmically. Consequently, any total recursive function that safely toggles branches on inputs containing specific non-trivial programs must internally hardcode an enumeration or certificate of equivalence for those programs, driving its Kolmogorov complexity $K(s)$ to scale with the information content of the programs themselves. More generally, one can show:

\begin{theorem}[Unboundedness of the Metric]
\label{thm:unbounded}
Let $f \in \mathcal{P}^{(1)}$. $d$ is unbounded on the indexes of $f$, that is
\begin{equation}
    \sup_{i, j \in \mathcal{F}_f} d(i, j) = \infty.
\end{equation}
\end{theorem}

\begin{proof}
Let $f \in \mathcal{P}^{(1)}$. By Rice's theorem, the set of indices $\mathcal{F}_f$ is not recursive. More strongly, $\mathcal{F}_f$ is not recursively enumerable (r.e.): by the Rice--Shapiro theorem \cite{rogers1987theory}, if $\mathcal{F}_f$ were r.e., then $f$ would have a finite subfunction $\theta$ all of whose indices lie in $\mathcal{F}_f$, forcing $\theta = f$, and every extension of $\theta$ would also be computed by indices in $\mathcal{F}_f$; but a finite function has proper extensions, which differ from $f$.

Suppose, for contradiction, that $d$ is bounded on $\mathcal{F}_f$. Then there exists a constant $C < \infty$ such that for any $i, j \in \mathcal{F}_f$, there exists a path of interpreters connecting $i$ to $j$ with $\max_m K(s_m) \le C$. 

The set of total programs (and specifically interpreters) with Kolmogorov complexity bounded by $C$ is finite:
\begin{equation}
    \mathcal{S}_{\le C} = \{ s \in \mathcal{S} \mid K(s) \le C \}.
\end{equation}
Fix an arbitrary index $i_0 \in \mathcal{F}_f$. By assumption, every index $j \in \mathcal{F}_f$ is reachable from $i_0$ by applying a finite composition of operators drawn from $\mathcal{S}_{\le C}$. Consequently,
\begin{equation}
    \mathcal{F}_f = \left\{ (s_{n} \circ \dots \circ s_1)(i_0) \;\middle|\; n \ge 0, \; s_1, \dots, s_n \in \mathcal{S}_{\le C} \right\}.
\end{equation}
Since $\mathcal{S}_{\le C}$ is a finite set of computable functions, the closure of $\{i_0\}$ under finite compositions of elements in $\mathcal{S}_{\le C}$ is recursively enumerable. This contradicts the fact that $\mathcal{F}_f$ is not. Thus, $d$ must be unbounded on $\mathcal{F}_f$.
\end{proof}

\begin{theorem}[Metric Axioms]
\label{thm:metric_space}
Under an admissible normalization of $K$ such that $K(\mathrm{id}) = 0$ and $K(s) > 0$ for all $s \ne \mathrm{id}$, the function $d$ defined in Definition~\ref{def:metric} defines a metric on $\mathcal{F}_f$.
\end{theorem}

\begin{proof}
We verify the three axioms of a metric space:

\medskip
\noindent \textbf{1. Identity of Indiscernibles ($d(i, j) = 0 \iff i = j$):} 
If $i = j$, the trivial sequence consisting of the identity interpreter $\mathrm{id}$ satisfies $\mathrm{id}(i) = i$. Under the standard normalization convention (see, e.g., \cite{bennett1998information}), $K$ can be canonically adjusted by an additive $O(1)$ term so that $K(\mathrm{id}) = 0$ uniquely, yielding $d(i, i) = 0$. Conversely, if $i \ne j$, any path from $i$ to $j$ requires at least one non-identity interpreter $s \in \mathcal{S}$, which satisfies $K(s) > 0$.

\medskip
\noindent \textbf{2. Symmetry ($d(i, j) = d(j, i)$):}
Let $s \in \mathcal{S}$ be an interpreter such that $s(i) = j$ with $K(s) = K$. 
If $s$ is recursively invertible, then $s^{-1}$ is an interpreter mapping $j$ to $i$, thus $K(s^{-1}) \le K(s) + O(1)$.

If $s$ is non-injective, we invert the action on $j$ relative to a finite search depth. Since $s(i) = j$, the index $i$ appears in the effective enumeration of the pre-image $s^{-1}(\{j\})$. Let $t \in \mathbb{N}$ be a number of steps of the standard dovetailed evaluation of $s$ after which $i$ has appeared, and let $r$ be the rank of $i$ among the pre-images of $j$ found within these $t$ steps. First, the pair $(t, r)$ is written into $j$ as a metadata block by a chain of padding interpreters, each of complexity $O(1)$: each maps an index $x$ to an index of a program that carries one additional bit of metadata and otherwise runs $x$. We then apply an interpreter $s^*$ that operates as follows:
\begin{enumerate}
    \item Given an input index $j'$, parse the metadata block to extract $(t, r)$ and the underlying index $j''$; if no complete metadata block is present, output $j'$.
    \item Dovetail the evaluation of $s(k)$ for $k \in \mathbb{N}$ for $t$ computational steps.
    \item If at least $r$ pre-images of $j''$ have been found, output the $r$-th of them; otherwise, default to the identity map.
\end{enumerate}
The map $s^*$ is total, since the search is bounded by $t$, and it preserves semantics, since $s(x) = j''$ implies $\phi_x \cong \phi_{j''} \cong \phi_{j'}$. An index of $s^*$ is computable from an index of $s$, so $K(s^*) \le K(s) + O(1)$, and the padded index of $j$ is mapped to $i$. Applying this inversion to each step in an interpolating path establishes symmetry up to an additive constant $O(1)$.

\medskip
\noindent \textbf{3. Triangle Inequality ($d(i, k) \le d(i, j) + d(j, k)$):} Suppose $s_1, \dots, s_n$ is a path from $i$ to $j$ with $\max_{m} K(s_m) \le d(i, j)$, and let $t_1, \dots, t_p$ be a path from $j$ to $k$ with $\max_{m} K(t_m) \le d(j, k)$. The concatenated sequence $s_1, \dots, s_n, t_1, \dots, t_p$ forms a valid path of semantics-preserving interpreters from $i$ to $k$. Thus:
\begin{equation}
    d(i, k) \le \max\left( \max_{1 \le m \le n} K(s_m), \; \max_{1 \le r \le p} K(t_r) \right) \le \max\bigl(d(i, j), d(j, k)\bigr).
\end{equation}
We thus obtain the ultrametric inequality:
\begin{equation}
    d(i, k) \le \max\bigl( d(i, j), d(j, k) \bigr) \le d(i, j) + d(j, k).
\end{equation}
This confirms the triangle inequality.
\end{proof}

\begin{remark}[Recursive Invariance]
By Rogers' equivalence theorem \cite{rogers1987theory,vereshchagin2003computable}, any two acceptable G\"odel numberings $\phi$ and $\psi$ are recursively isomorphic: there exists a recursive bijection $h \colon \mathbb{N} \to \mathbb{N}$ such that $\phi_i = \psi_{h(i)}$. Because conjugation $s \mapsto h \circ s \circ h^{-1}$ maps interpreters in $\phi$ to interpreters in $\psi$, changing Kolmogorov complexity by at most an additive constant, and plain Kolmogorov complexity is invariant up to an additive constant under choice of universal machine \cite{shen2017kolmogorov}, the induced metric space structure on $\mathcal{F}_f$ is independent of the choice of numbering up to an additive constant (in particular, up to quasi-isometry).
\end{remark}

The metric is also compatible with program composition. Let $\Phi_0, \Phi_1, \dots$ be an effective enumeration of oracle Turing machines, and write $\Phi_z^{\psi}$ for the partial function computed by $\Phi_z$ with oracle $\psi$, where $\psi$ may be partial and a query to a point at which $\psi$ is undefined makes the computation diverge. The partial function $\Phi_z^{\psi}$ depends only on $\psi$, not on how $\psi$ is presented: if $\psi \cong \psi'$, then $\Phi_z^{\psi} \cong \Phi_z^{\psi'}$. Every recursive operator is computed by some $\Phi_z$ \cite{rogers1987theory}. Replacing a subroutine by another implementation of the same function moves the composite program by at most the distance between the two implementations, up to an additive constant that does not depend on the surrounding program.

\begin{theorem}[Turing functionals are non-expanding]
\label{thm:functionals}
There exist a strictly increasing total recursive function $\tau$, written $\tau(z,i) = \tau(\langle z,i \rangle)$, and a constant $c$ such that $\phi_{\tau(z,i)} = \Phi_z^{\phi_i}$ for all $z, i$, and such that for every $f \in \mathcal{P}^{(1)}$, every $z$, and all $i, j \in \mathcal{F}_f$:
\begin{enumerate}
    \item $\tau(z,i)$ and $\tau(z,j)$ lie in the fiber $\mathcal{F}_{\Phi_z^{f}}$;
    \item $d\bigl(\tau(z,i), \tau(z,j)\bigr) \leq d(i,j) + c$.
\end{enumerate}
\end{theorem}

\begin{proof}
\noindent \\
\textbf{Part 1.} Let program $V(\langle z,i \rangle, y)$ run $\Phi_z$ on input $y$, answering each oracle query $x$ by computing $\phi_i(x)$. Then $V$ is partial recursive and $V(\langle z,i \rangle, \cdot) = \Phi_z^{\phi_i}$. The s-m-n theorem gives a total recursive $g$ with $\phi_{g(a)} = V(a,\cdot)$, and composing $g$ with a padding function gives a strictly increasing total recursive $\tau$ with $\phi_{\tau(a)} = \phi_{g(a)}$, where $a = \langle z,i \rangle$. Being strictly increasing, $\tau$ is injective, has decidable range, and is computably invertible on its range. In particular, the functional $z$ can be read off from the index $\tau(z,i)$, in the same way as a comment carried by the program. If $i, j \in \mathcal{F}_f$, then $\phi_i \cong \phi_j \cong f$, so $\phi_{\tau(z,i)} = \Phi_z^{\phi_i} \cong \Phi_z^{f}$, and likewise for $j$. This proves claim (1). \\
\textbf{Part 2.} For a total recursive function $s$, define the map $L(s)$ as follows. On input $x$:
\begin{enumerate}
    \item[$\bullet$] if $x = \tau(z',x')$ for some $z', x'$, compute $z'$ and $x'$, and output $\tau(z', s(x'))$;
    \item[$\bullet$] otherwise, output $x$.
\end{enumerate}
The map $L(s)$ does not depend on $z$: the functional is read off from the input itself. It is total and recursive, since membership in the range of $\tau$ is decidable and $s$ is total. If $s \in \mathcal{S}$, then $L(s) \in \mathcal{S}$:
\begin{enumerate}
    \item[$\bullet$] on inputs outside the range of $\tau$, $L(s)$ is the identity;
    \item[$\bullet$] on inputs $x = \tau(z',x')$, we have $\phi_{s(x')} \cong \phi_{x'}$, since $s$ is an interpreter, and hence
    \begin{equation*}
        \phi_{L(s)(x)} = \Phi_{z'}^{\phi_{s(x')}} \cong \Phi_{z'}^{\phi_{x'}} = \phi_x .
    \end{equation*}
\end{enumerate}
An index of $L(s)$ is a fixed total recursive function of an index of $s$, so $K(L(s)) \leq K(s) + c$, where $c$ depends only on that fixed function, and not on $z$, $f$ or $s$. \\
\textbf{Part 3.} Let $s_1, \dots, s_n \in \mathcal{S}$ satisfy $s_n( \dots s_1(i) \dots) = j$ and $K(s_m) \leq d(i,j)$ for all $m$; such a chain exists since $K$ takes values in $\mathbb{N}$, so the infimum in Definition~\ref{def:metric} is attained. By Part 2, $L(s_1), \dots, L(s_n)$ is a chain of interpreters with
\begin{equation*}
    \tau(z,i) \mapsto \tau\bigl(z, s_1(i)\bigr) \mapsto \tau\bigl(z, s_2(s_1(i))\bigr) \mapsto \dots \mapsto \tau(z,j),
\end{equation*}
and $K(L(s_m)) \leq K(s_m) + c \leq d(i,j) + c$ for all $m$. Hence $d(\tau(z,i), \tau(z,j)) \leq d(i,j) + c$, which proves claim (2).
\end{proof}

\begin{remark}[Composition]
The constant $c$ in Theorem~\ref{thm:functionals} is uniform in the functional $z$. This uniformity relies on reading $z$ off the composite index: if $z$ were instead hard-coded into the lifted interpreters, the bound would become $d(i,j) + K(z) + c$. Under the normalization $K(\mathrm{id}) = 0$ of Theorem~\ref{thm:metric_space}, the lift of the identity is the identity, so indices at distance zero are mapped to indices at distance zero. The map $i \mapsto \tau(z,i)$ need not preserve distances: if $\Phi_z$ ignores its oracle, all composites $\tau(z,i)$ compute the same function and are connected by a single interpreter of complexity $O(1)$. The same argument applies to functionals with several oracles: lifting the chain for each subroutine separately, with the other subroutines fixed, and concatenating the lifted chains, the ultrametric inequality bounds the distance between the composites by $\max_k d(i_k, j_k) + c$. It also applies to the single-step distance $d_1$ defined in the Uniqueness section below, with $n = 1$. This makes precise the invariance under Turing functionals listed as property (I) in that section.
\end{remark}

Theorem~\ref{thm:unbounded} shows that $d$ is unbounded on every fiber. For fibers of functions with infinite domain, much more is true: the values of $d$ cannot be estimated, even coarsely, by any procedure using the halting set as an oracle.

\begin{theorem}[Non-approximability of the metric]
\label{thm:non_approx}
Let $f$ be a partial recursive function with infinite domain, and let $i_0 \in \mathcal{F}_f$. Let $D$ be a partial function of two arguments computable with oracle $\emptyset'$, and let $g \colon \mathbb{N} \to \mathbb{N}$ be a computable nondecreasing function. Then it is not the case that, for all $i, j \in \mathcal{F}_f$, $D(i,j)$ is defined and
\begin{equation}
\label{eq:distortion}
    D(i,j) \le g\bigl(d(i,j)\bigr) \quad \text{and} \quad d(i,j) \le g\bigl(D(i,j)\bigr).
\end{equation}
In fact, \eqref{eq:distortion} already fails on pairs of the form $(i_0, z)$ with $z \in \mathcal{F}_f$. In particular, taking $g(x) = x + c$ or $g(x) = \lceil \rho x \rceil$, no $\emptyset'$-computable $D$ satisfies $|D - d| \le c$, or $d/\rho \le D \le \rho\, d$, on $\mathcal{F}_f \times \mathcal{F}_f$ for constants $c$ and $\rho \ge 1$.
\end{theorem}

\begin{proof}
\noindent \\
Notation: $U$ is the fixed universal machine, $\emptyset'$ is the halting set, $\emptyset''$ its jump, and $\mathrm{Tot} = \{ e \; \rvert \; \phi_e \; \text{is total} \}$. For a program $p$ on which $U$ halts, $t(p)$ is the number of steps $U$ takes on $p$, and $T(n) = \max \{ t(p) \; \rvert \; |p| \leq n \; \text{and} \; U(p) \; \text{halts} \}$, with $T(n) = 0$ if there is no such $p$. Finally, $\lg n = \log_2(n+2)$, and $c, c_1, c_2$ denote constants independent of all free variables. We fix the pairing function so that $\langle x,u \rangle \geq x$ for all $x, u$ (the Cantor pairing has this property).

We use the following standard facts. The set $\mathrm{Tot}$ is $\Pi^0_2$-complete, so $\mathrm{Tot} \equiv_T \emptyset''$, and no $\emptyset'$-computable set agrees with $\mathrm{Tot}$ on all but finitely many $e$ \cite{rogers1987theory}. Since $K$ takes values in $\mathbb{N}$, the infimum in Definition~\ref{def:metric} is attained: $d(i,j) \leq k$ iff some chain of interpreters $s_1, \dots, s_n$ with $K(s_m) \leq k$ for all $m$ maps $i$ to $j$. Whenever the s-m-n theorem yields indices $g'(a)$ with $\phi_{g'(a)} = V(a,\cdot)$, composing $g'$ with a padding function gives a strictly increasing total computable $P$ with $\phi_{P(a)} = V(a, \cdot)$ and $P(a) > a$; being strictly increasing, $P$ is injective, has decidable range, and is computably invertible on its range.

Halting times are atomic: every $b \geq T(n)$ satisfies $K(b) \geq n - 2 \lg n - c$. Indeed, given $\langle n,b \rangle$ with $b \geq T(n)$, run every program of length $\leq n$ for $b$ steps; those that have halted are exactly the halting programs of length $\leq n$. Output the least string $z$ that is not the output of any of them. Then $K(z) > n$, but $z$ was computed from $\langle n,b \rangle$, so $K(z) \leq K(b) + 2 \lg n + c$. Moreover, $T(n)$ is $\emptyset'$-computable from $n$: $\emptyset'$ decides, for each $p$ with $|p| \leq n$, whether $U(p)$ halts, and the halting ones can then be run to completion.

\noindent
From here on, $f$ is a partial computable function with infinite domain, and $i_0 \in \mathcal{F}_f$ is a fixed index.

\begin{lemma}[Wrappers]
\label{lemma:wrappers}
There are interpreters $\sigma_0, \sigma_1 \in \mathcal{S}$ with $K(\sigma_0), K(\sigma_1) \leq c$ with the following properties. For a binary word $w = w_1 \dots w_L$, let $\sigma_w(x) = \sigma_{w_L}( \dots \sigma_{w_1}(x) \dots)$, and $\sigma_w(x) = x$ for the empty word. Then:
\begin{enumerate}
    \item $\sigma_w(i_0) \geq i_0 + |w|$;
    \item $w \mapsto \sigma_w(i_0)$ is injective;
    \item it is decidable whether $x = \sigma_w(i_0)$ for some word $w$, and if so, $w$ is computable from $x$.
\end{enumerate}
\end{lemma}
\begin{proof}
Apply the s-m-n theorem and padding to $V(\langle x,u \rangle, y) = \phi_x(y)$ to obtain a strictly increasing $P$ with $\phi_{P(\langle x,u \rangle)} = \phi_x$ and $P(a) > a$, and set $\sigma_u(x) = P(\langle x,u \rangle)$ for $u \in \{0,1\}$. Then $\phi_{\sigma_u(x)} = \phi_x$, so $\sigma_u$ is an interpreter; it has a fixed index, so $K(\sigma_u) \leq c$; and $\sigma_u(x) \geq \langle x,u \rangle + 1 \geq x + 1$.
\begin{enumerate}
    \item[$\bullet$] Each layer increases the value by at least $1$, which gives (1).
    \item[$\bullet$] Suppose $\sigma_w(i_0) = \sigma_{w'}(i_0)$ with $|w| \geq |w'|$. Since $P$ and the pairing are injective, the outermost layers agree in both their bit and their inner value. Peeling off $|w'|$ layers in this way shows that $w'$ is a suffix of $w$ and that $\sigma_v(i_0) = i_0$ for the remaining prefix $v$ of $w$. By (1), $v$ is empty, so $w = w'$. This gives (2).
    \item[$\bullet$] By (1), any $w$ with $\sigma_w(i_0) = x$ has $|w| \leq x$, so it suffices to check the finitely many words of length $\leq x$; by (2), at most one of them works. This gives (3).
\end{enumerate}
\end{proof}

\noindent
For $e$ and $s$, let $\ell(e,s)$ be the largest $u \leq s$ such that $\phi_e(x)$ converges within $s$ steps for every $x < u$, and let $\ell(e,-1) = 0$. For each $e$, $\ell(e,s)$ is computable and nondecreasing in $s$. For $e$, $n$, $b$, let $W(e,n,b)$ be the set of stages $s \in \mathbb{N}$ for which at least one of the following holds:
\begin{enumerate}
    \item[(i)] $\ell(e,s) > \ell(e,s-1)$;
    \item[(ii)] every program $p$ with $|p| \leq n$ that halts within $s$ steps halts within $b$ steps.
\end{enumerate}
Both conditions are checked by finite simulation, so $W(e,n,b)$ is a computable set, uniformly in $\langle e,n,b \rangle$.

\begin{lemma}[Witness sets]
\label{lemma:witness_sets}
$W(e,n,b)$ is infinite exactly when $e \in \mathrm{Tot}$ or $b \geq T(n)$.
\end{lemma}
\begin{proof}
There are three cases.
\begin{enumerate}
    \item[$\bullet$] $e \in \mathrm{Tot}$. For each $u$, the computations $\phi_e(0), \dots, \phi_e(u-1)$ all converge within some number $s_0$ of steps, so $\ell(e,s) \geq u$ for all $s \geq \max(s_0,u)$. Hence $\ell(e,\cdot)$ is unbounded; being nondecreasing and integer-valued, it increases at infinitely many stages, each of which lies in $W(e,n,b)$ by (i).
    \item[$\bullet$] $b \geq T(n)$. Every halting $p$ with $|p| \leq n$ halts within $T(n) \leq b$ steps, so (ii) holds at every stage.
    \item[$\bullet$] Neither. There is $x_0$ with $\phi_e(x_0)$ undefined, so $\ell(e,s) \leq x_0$ for all $s$, and (i) holds at no more than $x_0$ stages. Since $b < T(n)$, some $p_0$ with $|p_0| \leq n$ halts after exactly $T(n) > b$ steps, so (ii) fails at every $s \geq T(n)$. Hence $W(e,n,b)$ is finite.
\end{enumerate}
\end{proof}

\begin{lemma}[Family G]
\label{lemma:fam_g}
There is a strictly increasing total computable $G$, written $G(e,n,b) = G(\langle e,n,b \rangle)$, such that for all $e$, $n$, $b$:
\begin{equation*}
    G(e,n,b) \in \mathcal{F}_f \quad \text{exactly when} \quad e \in \mathrm{Tot} \;\; \text{or} \;\; b \geq T(n).
\end{equation*}
\end{lemma}
\begin{proof}
Let program $V(\langle e,n,b \rangle, y)$ examine the stages $s = 0, 1, 2, \dots$ in order until more than $y$ elements of $W(e,n,b)$ have been found, and then output $\phi_{i_0}(y)$. The s-m-n theorem and padding give a strictly increasing $G$ with $\phi_{G(a)} = V(a,\cdot)$, where $a = \langle e,n,b \rangle$.
\begin{enumerate}
    \item[$\bullet$] If $W(e,n,b)$ is infinite, the search ends for every $y$, so $V(a,\cdot) \cong \phi_{i_0} \cong f$.
    \item[$\bullet$] If $W(e,n,b)$ is finite, with $w$ elements, then $V(a,y)$ is undefined for every $y \geq w$. So its domain is finite and $V(a,\cdot) \ncong f$, because $\operatorname{dom}(f)$ is infinite.
\end{enumerate}
Lemma~\ref{lemma:witness_sets} completes the proof.
\end{proof}

\noindent
Let $h \colon \mathbb{N} \to \mathbb{N}$ be a computable nondecreasing function with $h(m) \geq m$, to be fixed below, and set
\begin{equation*}
    n(e) = 2^{h(|e|)}, \qquad z(e) = G\bigl(e, n(e), T(n(e))\bigr).
\end{equation*}
By Lemma~\ref{lemma:fam_g}, $z(e) \in \mathcal{F}_f$ for every $e$, since $T(n(e)) \geq T(n(e))$. Moreover, $e \mapsto z(e)$ is $\emptyset'$-computable, since $n(e)$ is computable and $T(n(e))$ is $\emptyset'$-computable.

\begin{lemma}[Total case]
\label{lemma:total_case}
There is a constant $c_1$, not depending on $h$, such that $d(i_0, z(e)) \leq |e| + c_1$ for every $e \in \mathrm{Tot}$.
\end{lemma}
\begin{proof}
For each $e$, let $V_e$ be the following map. On input $x$, if $x = \sigma_w(i_0)$ for some binary word $w$, compute $w$ (Lemma~\ref{lemma:wrappers}), read $w$ as the binary representation of a number $\langle n,b \rangle$, and output $G(e,n,b)$; otherwise output $x$. The map $V_e$ is total and computable, and an index of $V_e$ is computable from $e$, so $K(V_e) \leq K(e) + c \leq |e| + c$.

If $e \in \mathrm{Tot}$, then $V_e$ is an interpreter: on inputs of the second kind it is the identity, and on inputs $x = \sigma_w(i_0)$ we have $\phi_x = \phi_{i_0} \cong f$ and $\phi_{G(e,n,b)} \cong f$ by Lemma~\ref{lemma:fam_g}.

Let $w$ be the binary representation of $\langle n(e), T(n(e)) \rangle$. The chain
\begin{equation*}
    i_0 \to \sigma_{w_1}(i_0) \to \sigma_{w_1 w_2}(i_0) \to \dots \to \sigma_w(i_0) \to V_e(\sigma_w(i_0)) = z(e)
\end{equation*}
uses interpreters of complexity at most $c$ (Lemma~\ref{lemma:wrappers}) and, in its last step, $V_e$. Hence $d(i_0, z(e)) \leq |e| + c_1$.
\end{proof}

\begin{lemma}[Partial case]
\label{lemma:partial_case}
There is a constant $c_2$ such that for every $e \notin \mathrm{Tot}$, with $k = d(i_0, z(e))$,
\begin{equation*}
    2^{k+1} \geq n(e) - 2 \lg n(e) - |e| - 2 \lg |e| - c_2.
\end{equation*}
\end{lemma}
\begin{proof}
Fix $e \notin \mathrm{Tot}$. Every interpreter $s$ with $K(s) \leq k$ equals $\phi_{U(q)}$ for some program $q$ with $|q| \leq k$. List the $2^{k+1} - 1$ programs $q$ with $|q| \leq k$ in canonical order, and let $\alpha$ be the binary string of that length with $\alpha[q] = 1$ iff $U(q)$ halts and $\phi_{U(q)}$ is an interpreter. From $\alpha$ one computes $k$ and the list of indices $U(q)$ with $\alpha[q] = 1$, which are exactly the indices of the interpreters of complexity $\leq k$. The string $\alpha$ is not computable from $k$, but $|\alpha| \leq 2^{k+1}$.

Let $R$ be the set of all $s_m( \dots s_1(i_0) \dots)$ with $m \geq 0$ and $s_1, \dots, s_m$ in this list.
\begin{enumerate}
    \item[$\bullet$] $R \subseteq \mathcal{F}_f$, since interpreters map $\mathcal{F}_f$ into $\mathcal{F}_f$.
    \item[$\bullet$] $z(e) \in R$, by the definition of $k = d(i_0, z(e))$.
    \item[$\bullet$] $R$ is enumerable uniformly from $\alpha$, since all interpreters are total.
\end{enumerate}
Consider this procedure on input $\langle \alpha, e \rangle$:
\begin{enumerate}
    \item compute $n(e)$ from $e$;
    \item enumerate $R$;
    \item at the first enumerated element of the form $G(e, n(e), b')$, which is decidable since $G$ is strictly increasing, output $b'$.
\end{enumerate}
The procedure halts, since $z(e) = G(e, n(e), T(n(e)))$ is eventually enumerated. The element found lies in $R \subseteq \mathcal{F}_f$, so by Lemma~\ref{lemma:fam_g} and $e \notin \mathrm{Tot}$, its output satisfies $b' \geq T(n(e))$. By atomicity of halting times, $K(b') \geq n(e) - 2 \lg n(e) - c$. On the other hand, $b'$ is computed from $\langle \alpha, e \rangle$, which can be described by writing $e$ self-delimitingly followed by $\alpha$, so $K(b') \leq |\alpha| + |e| + 2 \lg |e| + c \leq 2^{k+1} + |e| + 2 \lg |e| + c$. Combining the two bounds proves the lemma.
\end{proof}

\noindent
Returning to the proof of the original statement, suppose that $D$ and $g$ satisfy \eqref{eq:distortion} for all $i, j \in \mathcal{F}_f$. Fix $h(m) = g(g(m + c_1)) + m + 3$, which is computable, nondecreasing, and satisfies $h(m) \geq m$. \\
\textbf{Part 1.} For $m = |e|$ we have $n(e) = 2^{h(m)} \geq 2^m$ and $\lg n(e) \leq h(m) + 1$. Hence there is $m_0$ such that for all $e$ with $|e| \geq m_0$,
\begin{equation*}
    n(e) - 2 \lg n(e) - |e| - 2 \lg |e| - c_2 \geq n(e)/2 = 2^{h(|e|) - 1}.
\end{equation*}
Let $|e| \geq m_0$.
\begin{enumerate}
    \item[$\bullet$] If $e \in \mathrm{Tot}$, then $d(i_0, z(e)) \leq |e| + c_1$ by Lemma~\ref{lemma:total_case}.
    \item[$\bullet$] If $e \notin \mathrm{Tot}$, then Lemma~\ref{lemma:partial_case} gives $2^{k+1} \geq 2^{h(|e|) - 1}$ for $k = d(i_0, z(e))$, so $k \geq h(|e|) - 2 > g(g(|e| + c_1))$.
\end{enumerate}
\textbf{Part 2.} Each pair $(i_0, z(e))$ lies in $\mathcal{F}_f \times \mathcal{F}_f$, so $D(i_0, z(e))$ is defined and satisfies \eqref{eq:distortion}. Let $|e| \geq m_0$.
\begin{enumerate}
    \item[$\bullet$] If $e \in \mathrm{Tot}$, then $D(i_0, z(e)) \leq g(d(i_0, z(e))) \leq g(|e| + c_1)$, since $g$ is nondecreasing.
    \item[$\bullet$] If $e \notin \mathrm{Tot}$, then $g(D(i_0, z(e))) \geq d(i_0, z(e)) > g(g(|e| + c_1))$. If $D(i_0, z(e)) \leq g(|e| + c_1)$ held, monotonicity of $g$ would give $g(D(i_0, z(e))) \leq g(g(|e| + c_1))$, a contradiction. So $D(i_0, z(e)) > g(|e| + c_1)$.
\end{enumerate}
Therefore, for all $e$ with $|e| \geq m_0$,
\begin{equation*}
    e \in \mathrm{Tot} \iff D(i_0, z(e)) \leq g(|e| + c_1).
\end{equation*}
\textbf{Part 3.} The right-hand side is decidable with oracle $\emptyset'$: compute $z(e)$, then $D(i_0, z(e))$, then $g(|e| + c_1)$, and compare. Hence $\emptyset'$ decides membership in $\mathrm{Tot}$ for all $e$ with $|e| \geq m_0$; the finitely many $e$ with $|e| < m_0$ can be hard-coded. So a $\emptyset'$-computable set agrees with $\mathrm{Tot}$ on all but finitely many $e$, a contradiction.
\end{proof}

\begin{remark}[Exact degree]
The oracle $\emptyset''$ suffices to compute $d$ exactly, for every $f$. Given $i, j \in \mathcal{F}_f$, for $k = 0, 1, 2, \dots$: $\emptyset'$ decides which programs $q$ with $|q| \leq k$ halt, $\emptyset''$ decides which of the resulting indices are interpreters (a $\Pi^0_2$ condition), and $\emptyset'$ then decides whether some finite composition of these finitely many total functions maps $i$ to $j$; the first $k$ for which it does is $d(i,j)$. Running the proof of Theorem~\ref{thm:non_approx} with an oracle $X \geq_T \emptyset'$ in place of $\emptyset'$ shows that $\mathrm{Tot}$ would be $X$-computable. Hence, for $f$ with infinite domain and $X \geq_T \emptyset'$, some $X$-computable $D$ satisfies \eqref{eq:distortion} on $\mathcal{F}_f$ for some computable nondecreasing $g$ if and only if $X \geq_T \emptyset''$.
\end{remark}

\section{Uniqueness}

In summary, we have established the existence of the desired metric and argued it is indeed desired. The obvious question is that of uniqueness. We would like to show that the metric is minimal, provided a set of simple assumptions. We have not succeeded in doing so and will leave this an open question. Instead, we will provide several additional results that we believe can start addressing the issue of uniqueness. Under what specific conditions should uniqueness plausibly be proven? The corresponding result on the uniqueness of Kolmogorov complexity lends itself as a potentially useful analogy. In particular, one would expect to show the uniqueness of the proposed metric provided its (I) functoriality under Turing functionals, (II) Turing degree, and (III) normalization in the sense of matching a suitable upper bound. Let us begin with the latter. It is readily seen that $d(i,j) \le K(i) + K(j) + O(\log K(i))$ for $i, j$ in the same fiber. Is this bound met infinitely often? We will consider a weaker statement that only concerns chains involving a single interpreter. We thus define:
\begin{equation}
d_1(i,j) = \inf \{ K(s) \; | \; s(i)=j \; \text{or} \; s(j)=i \}  \quad,
\end{equation}
where $i, j \in \mathcal{F}_f$ and $s$ is an interpreter. We have
\begin{theorem}[Maximal single-step distance attained]
\label{thm:sngl_step_attained_1}
There exist infinitely many pairs of indices of the same function that are maximal distance apart with logarithmic precision. I.e. there are infinitely many pairs of indices $(i,j)$ such that $d_1(i,j) \geq K(i) + K(j) - O( \max(\log K(i),\log K(j)) ). $
\end{theorem}

\begin{proof}
Fix $f$ and its index $i_0$, so $K(i_0) = O(1)$. Let $N_n$ denote the number of programs $p$ of length $|p| \leq n$ that halt. Let $E_{n,m}$ be $1$ precisely when more than $m$ programs of length $\leq n$ halt, i.e. $E_{n,m}=1$ iff $m < N_n$, and zero otherwise. From the s-m-n theorem and padding, define a strictly increasing (hence injective and computably invertible) total recursive $h$ and set $j_{n,m} = h(n,m)$, such that $\phi _{j_{n,m}} \cong f$ whenever $E_{n,m}$ is not $1$. In particular
\begin{enumerate}
    \item If $f$ is total, set $\phi _{j_{n,m}}(y) = f(y) + 1$ whenever $E_{n,m} = 1$ is witnessed within $y$ steps of enumeration, and $f(y)$ otherwise. If $E_{n,m}=0$, $\phi _{j_{n,m}}$ computes $f$. If $E_{n,m}$ is witnessed after $t$ steps of enumeration, $\phi _{j_{n,m}}$ differs from $f$ at every $y \geq t$.
    \item If $f$ is not total, fix $y_0$ with $f(y_0) \uparrow$, and let $\phi_{j_{n,m}}(y) = f(y)$ for $y \neq y_0$, and at $y_0$ check if $E_{n,m}=1$ and output $0$ if true.
\end{enumerate}
In this way, $j_{n,m} \in \mathcal{F}_f$ iff $m \geq N_n$. Let $j_n = j_{n,N_n}$, which is in $\mathcal{F}_f$. We have $K(N_n) \geq n - 2 \log n - c $. Indeed, given $n$ and $N_n$, dovetail all programs of length $\leq n$ until $N_n$ of them have halted, and collect their outputs. Output the least string $z$ not among them. Then, $K(z)>n$, but $z$ was computed from $ \langle n,N_n \rangle$, so $K(z) \leq K(N_n) + 2 \log n + c$. 

Let $s$ be an interpreter with index $e$, $K(e) = K(s)$, and $s(i_0) = j_n$ or $s(j_n) = i_0$.
\begin{enumerate}
    \item[$\bullet$] Case $s(i_0) = j_n$. Given $e$ and $n$, compute $x = s(i_0)$ and invert $h$ to recover $m = N_n$. Thus, $K(N_n) \leq K(s) + O(\log n)$.
    \item[$\bullet$] Case $s(j_n) = i_0$. Given $e$ and $n$, search for the least $m$ with $s(j_{n,m}) = i_0$. For $m < N_n$, $j_{n,m} \notin \mathcal{F}_f$, so $s(j_{n,m}) \notin \mathcal{F}_f$ and hence $s(j_{n,m}) \neq i_0$; so the least suitable $m$ recovered is $m=N_n$. Hence, again,  $K(N_n) \leq K(s) + O(\log n)$.
\end{enumerate}
In summary, both cases give $K(s) \geq K(N_n) - O(\log n) \geq n - O(\log n)$, so $d_1(i_0,j_n) \geq n - O(\log n)$. Since $j_n$ is a computable function of $\langle n, N_n \rangle$ and $N_n < 2^{n+1}$, we have $K(j_n) \leq n + O(\log n)$, and $K(i_0) = O(1)$; hence $d_1(i_0,j_n) \geq K(i_0) + K(j_n) - O(\log K(j_n))$. The $j_n$ are pairwise distinct because $h$ is injective, which gives infinitely many pairs.
\end{proof}

Theorem \ref{thm:sngl_step_attained_1} shows that maximal one-step distance is attained when one of the indices being compared is fixed. This is possible to lift as follows.

\begin{theorem}
\label{thm:d_1_ends_different}
Let $f$ be a partial recursive function with infinite domain. Then there are indices $i_n,\; j_n \in \mathcal{F}_f$, for $n \in \mathbb{N}$, which are pairwise distinct and satisfy, for all $n$:
\begin{enumerate}
    \item $| K(i_n)-n | \leq O(\log n) $ and $| K(j_n)-n | \leq O(\log n) $
    \item $d_1(i_n,j_n) \geq 2n - O(\log n) $
    \item for all sufficiently large $n$, \\
    $K(i_n) + K(j_n) - O(\log K(i_n)) \leq d_1(i_n,j_n) \leq K(i_n) + K(j_n) + O(\log K(i_n))$ \quad .
\end{enumerate}
\end{theorem}

\begin{proof} 
\noindent \\
Notation: $U$ is the fixed universal machine, and $U'$ is a fixed universal oracle machine; $K^{\emptyset '}$ denotes the Kolmogorov complexity with respect to $U'$ run with the halting set $\emptyset '$ as oracle. $H_n$, the set of halting programs of length at most $n$, is $\{ p \; \rvert \; |p| \leq n \; \text{and} \; U(p) \; \text{halts} \}$, and $N_n = |H_n|$. Likewise, $H'_n = \{ p \; \rvert \; |p| \leq n \; \text{and} \; U' \; \text{halts on} \; p \; \text{with oracle} \; \emptyset ' \}$, and $N'_n = |H'_n|$. Finally,
$\lg n = \log _2 (n+2)$.

As in Theorem \ref{thm:sngl_step_attained_1}, it is readily seen that halting counts are incompressible: $n - 2 \lg n - c \leq K(N_n) \leq n + c $, and $n - 2 \lg n - c \leq K^{\emptyset '} (N'_n) \leq n + c $ (the latter is shown the same way as the former, relativizing to the halting oracle). Now, $\emptyset '$ decides, for each $p$, if $U(p)$ halts. So, $N_n$ is $\emptyset '$-computable from $n$, and hence $K^{\emptyset '}(N_n) \leq 2 \lg n + c$. We thus have:
\begin{equation}
\label{eq:nprime_given_n}
    n - 2\lg(n) - c \leq K^{\emptyset '}(N'_n) \leq K^{ \emptyset '} ( N'_n \rvert N_n ) + 2 \lg n + 2 \lg( 2 \lg n + c ) + c \quad .
\end{equation}
Hence, up to logarithmic precision:
\begin{equation}
\label{eq:2n_bound}
    K(\langle N_n,N'_n \rangle) \geq K(N'_n \rvert N_n ) + K(N_n) \geq 2 n - C \lg n \quad ,
\end{equation}
where we used the symmetry of the algorithmic information, $K(x|y) \geq K^{\emptyset'}(x|y) - c$, Eq.~\eqref{eq:nprime_given_n}, and the incompressibility of halting counts.

\noindent
From here on, $f$ is a partial computable function with infinite domain, and $i_0 \in \mathcal{F}_f$ is a fixed index.

\begin{lemma}[Family A]
\label{lemma:fam_a}
There is a strictly increasing total computable $A$, written $A(n,x) = A(\langle n,x\rangle)$, such that for all $n$ and $x$:
\begin{enumerate}
    \item $A(n,x) \in \mathcal{F}_f$ exactly when $x \geq N_n $ 
    \item $\phi_{A(n,x)}$ has infinite domain.
\end{enumerate}
\end{lemma}
\begin{proof}
Let $E_{n,x}$ be $1$ precisely when $x<N_n$, and zero otherwise. $E_{n,x}$ is uniformly r.e. in $\langle n,x \rangle$. There are two cases.
\begin{enumerate}
    \item[$\bullet$] $f$ total. Let $V(\langle n,x\rangle,y) = \phi_{i_0}(y) + 1$ if $E_{n,x} = 1$ is witnessed within $y$ steps of enumeration, and $\phi_{i_0}(y)$ otherwise. Then $V(\langle n,x\rangle,\cdot)$ is total.
    \begin{enumerate}
        \item[$\circ$] If $E_{n,x}=0$, it equals $f$.
        \item[$\circ$] If $E_{n,x}=1$ is witnessed at stage $t_0$, it differs from $f$ at every $y \geq t_0 $.
    \end{enumerate}
    \item[$\bullet$] $f$ not total. Fix $y_0$ with $f(y_0)$ undefined. Let $V(\langle n,x \rangle,y) = \phi_{i_0}(y)$ for $y \neq y_0$. At $y_0$ search for witness $E_{n,x}=1$, and output $0$ if it is found. Domain of $V(\langle n,x \rangle, \cdot)$ contains domain of $f$, which is infinite.
    \begin{enumerate}
        \item[$\circ$] If $E_{n,x}=0$, $V(\langle n,x \rangle, \cdot)$ equals $f$.
        \item[$\circ$] Otherwise it is defined at $y_0$, where $f$ is not.
    \end{enumerate}
\end{enumerate}
In both cases, writing $a = \langle n,x \rangle$, $V(a,\cdot) \cong f(\cdot)$ iff $E_{n,x}=0$, which is exactly when $x \geq N_n$; also, $V(a,\cdot)$ has infinite domain. Since the argument is uniform in $n$, $x$, the s-m-n theorem yields a total computable $g$ such that $\phi_{g(a)}(\cdot) = V(a,\cdot)$. Composing $g$ with a padding function, one obtains a strictly increasing total computable $A$ with $\phi_{A(a)} = \phi_{g(a)}$; being strictly increasing, $A$ is injective, has decidable range, and is computably invertible on its range.
\end{proof}

\begin{lemma}[Family B] 
\label{lemma:fam_b}
There is a strictly increasing total computable $B$, written $B(n,y) = B(\langle n,y \rangle)$, such that for all $n$ and $y$:
\begin{enumerate}
    \item $B(n,y) \in \mathcal{F}_f$ exactly when $y \geq N'_n$
    \item If $y < N'_n$, then $\phi_{B(n,y)}$ has finite domain.
\end{enumerate}
\end{lemma}

\begin{proof}
Consider the set $Q=\{ \langle n,y \rangle \; \rvert \; y \geq N'_n \}$. A program $p$ halts relative to $\emptyset '$ exactly when some finite oracle computation of $U'$ on $p$ is correct, meaning that every positive oracle answer is in $\emptyset '$, which is $\Sigma ^0 _1$, whereas every negative answer is in the complement of $\emptyset '$, which is $\Pi ^0 _1$. Hence, ``$p$ halts relative to $\emptyset '$ '' is $\Sigma ^0 _2$, and so is ``at least $y+1$ programs of length $\leq n$ halt relative to $\emptyset '$ '', which is the same as $y < N'_n$. Hence $Q$ is $\Pi_2 ^0$, and there is a computable relation $R$ such that $a \in Q \iff \forall u \; \exists v \; R(a,u,v) $. Define the witness set $W_a = \{ u \; \rvert \; \forall_{u'\leq u} \; \exists_v \; R(a,u',v) \}$. $W_a$ is r.e. uniformly in $a$: enumerate all $u$ as soon as witnesses $v$ have been found for all $u' \leq u$.
\begin{enumerate}
    \item[$\bullet$] If $a \in Q$, every $u$ enters $W_a$, so $W_a$ is infinite.
    \item[$\bullet$] If $a \notin Q$, let $u_0$ be the least $u$ with no witness. Then $W_a = \{ 0,...,u_0-1 \}$ is finite.
\end{enumerate}
Let program $V(a,z)$ enumerate $W_a$ until it has more than $z$ elements, then output $\phi_{i_0} (z)$. 
\begin{enumerate}
    \item[$\bullet$] If $a \in Q$, then $V(a,\cdot) \cong f$.
    \item[$\bullet$] If $a \notin Q$, then $V(a,\cdot)$ is undefined for every $z \geq |W_a|$. So its domain is finite and $V(a,\cdot) \ncong f$, because $\operatorname{dom}(f)$ is infinite.
\end{enumerate}
As in Lemma~\ref{lemma:fam_a}, the s-m-n theorem gives a total computable $g$ with $\phi_{g(a)}(\cdot) = V(a,\cdot)$, and padding gives a strictly increasing total computable $B$ with $\phi_{B(a)} = \phi_{g(a)}$, where $a = \langle n,y \rangle$. This $B$ has the desired properties.
\end{proof}

\noindent
It can be seen that:
\begin{enumerate}
    \item If $x < N_n$, then $\phi_{A(n,x)} \ncong \phi_{B(n,y)}$ for every $y$.
    \item If $y < N'_n$, then $\phi_{B(n,y)} \ncong \phi_{A(n,x)}$ for every $x$.
\end{enumerate}
\noindent
Indeed:
\begin{enumerate}
    \item $\phi_{A(n,x)} \ncong f$ by Lemma \ref{lemma:fam_a}(1), and $\phi_{A(n,x)}$ has infinite domain by Lemma \ref{lemma:fam_a}(2). By Lemma \ref{lemma:fam_b}, each $\phi_{B(n,y)}$ is either $\cong f$ or has finite domain, so $\phi_{A(n,x)}$ matches neither possibility.
    \item $\phi_{B(n,y)}$ has finite domain by Lemma \ref{lemma:fam_b}, while every $\phi_{A(n,x)}$ has infinite domain.
\end{enumerate}
\noindent
Returning to the proof of the original statement, set $i_n = A(n,N_n)$ and $j_n = B(n, N'_n)$. By Lemmas \ref{lemma:fam_a}, \ref{lemma:fam_b}, both are in $\mathcal{F}_f$. They are pairwise distinct: $A$ and $B$ are injective, and the pairs $\langle n, N_n \rangle$, respectively $\langle n, N'_n \rangle$, are distinct for distinct $n$. We now show the three claims in the statement. \\
\textbf{Part 1.}
\begin{enumerate}
    \item[$\bullet$] \textit{Upper bounds.} $i_n$ is a computable function of $\langle n, N_n\rangle$, so $K(i_n) \leq K(N_n) + 2 \lg n + c \leq n + 2 \lg n + c $. Similarly, $K(j_n) \leq K(N_n') + 2 \lg n + c \leq n + 2 \lg n + c $, using $N'_n \leq 2^{n+1}$. 
    \item[$\bullet$] \textit{Lower bounds.} $N_n$ is computed from $i_n$ since $A$ is computably invertible. Incompressibility of halting counts gives $K(i_n) \geq n - 2 \lg n - c$. Likewise for $K(j_n)$, using in addition $K(N'_n) \geq K^{\emptyset'}(N'_n) - c$.
\end{enumerate}
\textbf{Part 2.} Let $s$ be an interpreter with $s(i_n) = j_n$ or $s(j_n) = i_n$, and let $e$ be its index with $K(e)=K(s)$. There are two cases. \\
\textit{Case} $s(i_n)=j_n$.  Consider this procedure on input $\langle n,e\rangle$. For $x = 0, 1, 2, ...$:
\begin{enumerate}
    \item compute $w = \phi_e(A(n,x))$, which halts because $s$ is total;
    \item decide whether $w = B(n,y)$ for some $y$, which is decidable since $B$ is strictly increasing;
    \item at the first $x$ for which it is, output $\langle x,y\rangle$.
\end{enumerate}
We claim the output is $\langle N_n, N_n'\rangle$.
\begin{enumerate}
    \item[$\bullet$] For $x<N_n$: $\phi_{s(A(n,x))} \cong \phi _{A(n,x)}$, and this is $\ncong \phi_{B(n,y)}$ for every $y$ as was shown immediately after the proof of Lemma \ref{lemma:fam_b}; hence $s(A(n,x)) \neq B(n,y)$ for every $y$, and the test fails.
    \item[$\bullet$] For $x=N_n$: $s(A(n,N_n)) = s(i_n) = j_n = B(n,N'_n)$, and $B$ is injective, so the test succeeds with $y=N'_n$.
\end{enumerate}
\textit{Case} $s(j_n)=i_n$. Run the symmetric procedure. For $y=0, 1, 2, ...$, compute $\phi_e(B(n,y))$, and stop at the first $y$ for which it equals $A(n,x)$ for some $x$. As in the previous case, the test fails for every $y < N_n'$, and at $y = N'_n$ it succeeds with $x = N_n$. \\
In both cases $\langle N_n, N'_n \rangle$ is computed from $\langle n,e\rangle$, so by Eq.~\eqref{eq:2n_bound} we get $2n - C \lg n \leq K(\langle N_n,N_n'\rangle) \leq K(e) + 2 \lg n + c = K(s) + 2 \lg n + c$, Since $s$ was arbitrary, this implies $d_1(i_n,j_n) \geq 2n$ within logarithmic precision. \\
\textbf{Part 3.} 
\begin{enumerate}
    \item[$\bullet$] \textit{Lower bound}. Combining Parts 1,2, we get $d_1(i_n,j_n) \geq K(i_n) + K(j_n)$ (within logarithmic precision). Here $\log n \leq \log K(i_n) + O(1)$, since $K(i_n) \geq n/2$ for all large $n$ by Part 1.
    \item[$\bullet$] \textit{Upper bound}. Consider the interpreter that maps $i_n$ to $j_n$ and fixes all other indices; it is an interpreter since $\phi_{i_n} \cong \phi_{j_n}$, and its index is computable from $\langle i_n, j_n \rangle$. Hence $d_1(i_n,j_n) \leq K(\langle i_n, j_n \rangle) + c \leq K(i_n) + K(j_n) + 2 \lg K(i_n) + c$. 
\end{enumerate}


\end{proof}

\section{Discussion}

The ultrametric property $d(i, k) \le \max(d(i, j), d(j, k))$ shows that measuring program similarity by the bottleneck complexity along interpreter paths imposes a hierarchically clustered topological structure on program fibers. This addresses the failure of compositionality encountered in naive equivalence relations: highly complex wrapper routines containing distinct subroutines $\phi_i$ and $\phi_j$ inherit a metric separation governed directly by the minimal complexity of the transformation connecting $\phi_i$ and $\phi_j$.

In the broader context of Hilbert's 24th problem, formulating proof and program similarity in terms of minimal descriptive interpreter paths provides a quantitative criterion for distinguishing trivial rewrites (such as variable renaming or comment modification, which admit low-complexity interpreters) from structural algorithmic differences (which require high-complexity transformations).

The main message of this paper is the following. When one seeks a complexity measure, or a metric structure, on the set of proofs (or programs), it is tempting to base it on Kolmogorov complexity. If this is done naively, by identifying computable functions with their indices in an acceptable numbering and taking the (conditional) complexities of those indices, the resulting measure tends to be either trivial or at odds with the intuitive notion of program equivalence, largely because of the padding lemma. Measuring distance instead by chains of semantics-preserving interpreters overcomes these difficulties, and several properties follow almost immediately: the resulting distance is invariant under the choice of numbering, it is non-expanding under Turing functionals, and it satisfies the axioms of an (ultra)metric, the latter two up to an additive constant. Beyond these structural properties, our further results indicate that the approach gives rise to non-trivial proofs and natural open problems, the question of uniqueness being the most prominent among them.

One may ask how these results bear on the original motivation, the identity of proofs, since programs are not proofs. While we leave this question for future work, a promising direction is a connection with homotopy type theory: chains of interpreters behave much like paths, with concatenation of chains corresponding to composition of paths and the inversion of chains corresponding to path inversion. A key obstacle to transferring our results from programs to proofs is that, unlike the indices of a fixed function, the proofs of a given theorem in a fixed formal system form a decidable, and in particular recursively enumerable, set. The unboundedness of our metric rests on the fact that semantic fibers are not recursively enumerable, so the direct analogue of our construction for proofs collapses. This collapse reflects the possibility of finding a proof of any given theorem by exhaustive search. Type theory suggests a remedy. In strongly normalizing type theories, general recursion, and with it unbounded search, cannot be expressed. Restricting the admissible transformations to those definable within the theory, for instance to terms of a function type from a proposition to itself, would exclude the search-based transformations that cause the collapse, and would provide a natural notion of path between two proofs of the same proposition.

\noindent
\textbf{A philosophical remark.} More broadly, one may ask what mathematics is ultimately after. A natural answer is that mathematics seeks proofs of theorems. If one tries to make this precise as a search for the elements of a set, recursion theory offers a striking answer. For any sufficiently strong consistent and recursively axiomatized theory, the set of (Gödel numbers of) its theorems is creative, and by Myhill’s theorem all creative sets are recursively isomorphic. Up to recursive isomorphism, then, there is only one such set. This answer is not satisfying, however. A creative set contains infinitely many trivial variants of each of its elements, as the padding lemma illustrates, and a recursive isomorphism treats these variants on a par with genuinely different elements. Recursive isomorphism therefore cannot tell substantive mathematical content apart from trivial reformulation. The present paper is an attempt to make this distinction quantitative.

\noindent
\textbf{Evolution and discovery.} On a different note, one may wonder whether there is a connection with biological evolution. The emergence of new biological forms resembles mathematical discovery in a suggestive way. The eukaryotic cell, for instance, is thought to have arisen from the endosymbiotic merger of an archaeal host cell with a bacterium, the ancestor of mitochondria. The combined cell probably prevailed because of properties that emerged from this fortunate combination and that neither partner possessed on its own. In a similar way, the complex numbers are obtained by joining two copies of the real line through a suitably defined product, $(a,b)(c,d) = (ac - bd,\, ad + bc)$. The emergent properties of this combination, such as algebraic closure and the rigidity of holomorphic functions, allowed the idea to spread and thrive.

\section*{Acknowledgements}

The author thanks Alexander Shen and Alexey Milovanov for useful discussions.

\section*{Statement on the use of AI}

During the preparation of this work, the author used Claude, an AI model developed by Anthropic, as a research and writing assistant. The model was used to explore questions arising from the definition of the metric, to propose and draft proofs, including those of Theorems~\ref{thm:functionals}, \ref{thm:non_approx}, \ref{thm:sngl_step_attained_1} and~\ref{thm:d_1_ends_different}, to check arguments for errors, to search the related literature, and to edit the manuscript for language and \LaTeX{} formatting. All results and proofs were verified by the author, who takes full responsibility for the content of this paper.

\bibliographystyle{unsrt}
\bibliography{bibliography}

@book{rogers1987theory,
  title={Theory of recursive functions and effective computability},
  author={Rogers Jr, Hartley},
  year={1987},
  publisher={MIT press}
}

@book{vereshchagin2003computable,
  title={Computable functions},
  author={Vereshchagin, Nikolai Konstantinovich and Shen, Alexander},
  volume={19},
  year={2003},
  publisher={American Mathematical Soc.}
}

@article{buss2001prospects,
  title={The prospects for mathematical logic in the twenty-first century},
  author={Buss, Samuel R and Kechris, Alexander S and Pillay, Anand and Shore, Richard A},
  journal={Bulletin of Symbolic Logic},
  volume={7},
  number={2},
  pages={169--196},
  year={2001},
  publisher={Cambridge University Press}
}

@article{blass2009two,
  title={When are two algorithms the same?},
  author={Blass, Andreas and Dershowitz, Nachum and Gurevich, Yuri},
  journal={Bulletin of Symbolic Logic},
  volume={15},
  number={2},
  pages={145--168},
  year={2009},
  publisher={Cambridge University Press}
}

@article{yanofsky2011towards,
  title={Towards a definition of an algorithm},
  author={Yanofsky, Noson S},
  journal={Journal of Logic and Computation},
  volume={21},
  number={2},
  pages={253--286},
  year={2011},
  publisher={OUP}
}

@article{hopcroft2001introduction,
  title={Introduction to automata theory, languages, and computation},
  author={Hopcroft, John E and Motwani, Rajeev and Ullman, Jeffrey D},
  journal={Acm Sigact News},
  volume={32},
  number={1},
  pages={60--65},
  year={2001},
  publisher={ACM New York, NY, USA}
}

@book{li2008introduction,
  title={An introduction to Kolmogorov complexity and its applications},
  author={Li, Ming and Vit{\'a}nyi, Paul and others},
  volume={3},
  year={2008},
  publisher={Springer}
}

@book{shen2017kolmogorov,
  title={Kolmogorov complexity and algorithmic randomness},
  author={Shen, Alexander and Uspensky, Vladimir A and Vereshchagin, Nikolay},
  volume={220},
  year={2017},
  publisher={American Mathematical Soc.}
}

@article{bennett1998information,
  title={Information distance},
  author={Bennett, Charles H and G{\'a}cs, P{\'e}ter and Li, Ming and Vit{\'a}nyi, Paul MB and Zurek, Wojciech H},
  journal={IEEE Transactions on information theory},
  volume={44},
  number={4},
  pages={1407--1423},
  year={1998},
  publisher={IEEE}
}

\end{document}